\documentclass[journal=nalefd,manuscript=letter,layout=twocolumn]{achemso}

\usepackage[version=3]{mhchem} 
\usepackage[T1]{fontenc} 
\usepackage[usenames,dvipsnames]{xcolor}


\title{Single Graphene Quantum dots in a single molecular nanoparticle}
\vspace{0.3cm} \vspace{0.3cm}
\author{Huynh Thanh Trung}
\affiliation[LUMIN]
{LuMIn, Universit\'e Paris-Saclay, ENS Paris Saclay, CNRS, Centrale Supelec, 91405 Orsay Cedex, France}

\author{Suman Sarkar}
\affiliation[LUMIN]
{LuMIn, Universit\'e Paris-Saclay, ENS Paris Saclay, CNRS, Centrale Supelec, 91405 Orsay Cedex, France}

\author{Oc\'eane Capelle}
\affiliation[LUMIN]
{LuMIn, Universit\'e Paris-Saclay, ENS Paris Saclay, CNRS, Centrale Supelec, 91405 Orsay Cedex, France}

\author{Sébastien Quistrebert}
\affiliation[LUMIN]
{LuMIn, Universit\'e Paris-Saclay, ENS Paris Saclay, CNRS, Centrale Supelec, 91405 Orsay Cedex, France}

\author{Frédéric fossard}
\affiliation[LEM]
{LEM, ONERA, CNRS, 92322 Chatillon, France}

\author{Daniel Medina-Lopez}
\affiliation[CEA]{Université Paris-Saclay, CEA, CNRS, NIMBE, LICSEN, 91191 Gif-sur-Yvette, France}

\author{Cynthia Banga-Kpako}
\affiliation[CEA]{Université Paris-Saclay, CEA, CNRS, NIMBE, LICSEN, 91191 Gif-sur-Yvette, France}

\author{Elsa Cassette}
\affiliation[LUMIN]
{LuMIn, Universit\'e Paris-Saclay, ENS Paris Saclay, CNRS, Centrale Supelec, 91405 Orsay Cedex, France}

\author{St\'ephane Campidelli}
\affiliation[CEA]{Université Paris-Saclay, CEA, CNRS, NIMBE, LICSEN, 91191 Gif-sur-Yvette, France}

\author{Jean-S\'ebastien Lauret}
\affiliation[LUMIN]
{LuMIn, Universit\'e Paris-Saclay, ENS Paris Saclay, CNRS, Centrale Supelec, 91405 Orsay Cedex, France}
\email{lauret@ens-paris-saclay.fr}
\phone{+33-1-69352132}

\begin{document}

\begin{tocentry}

\includegraphics[scale=.18]{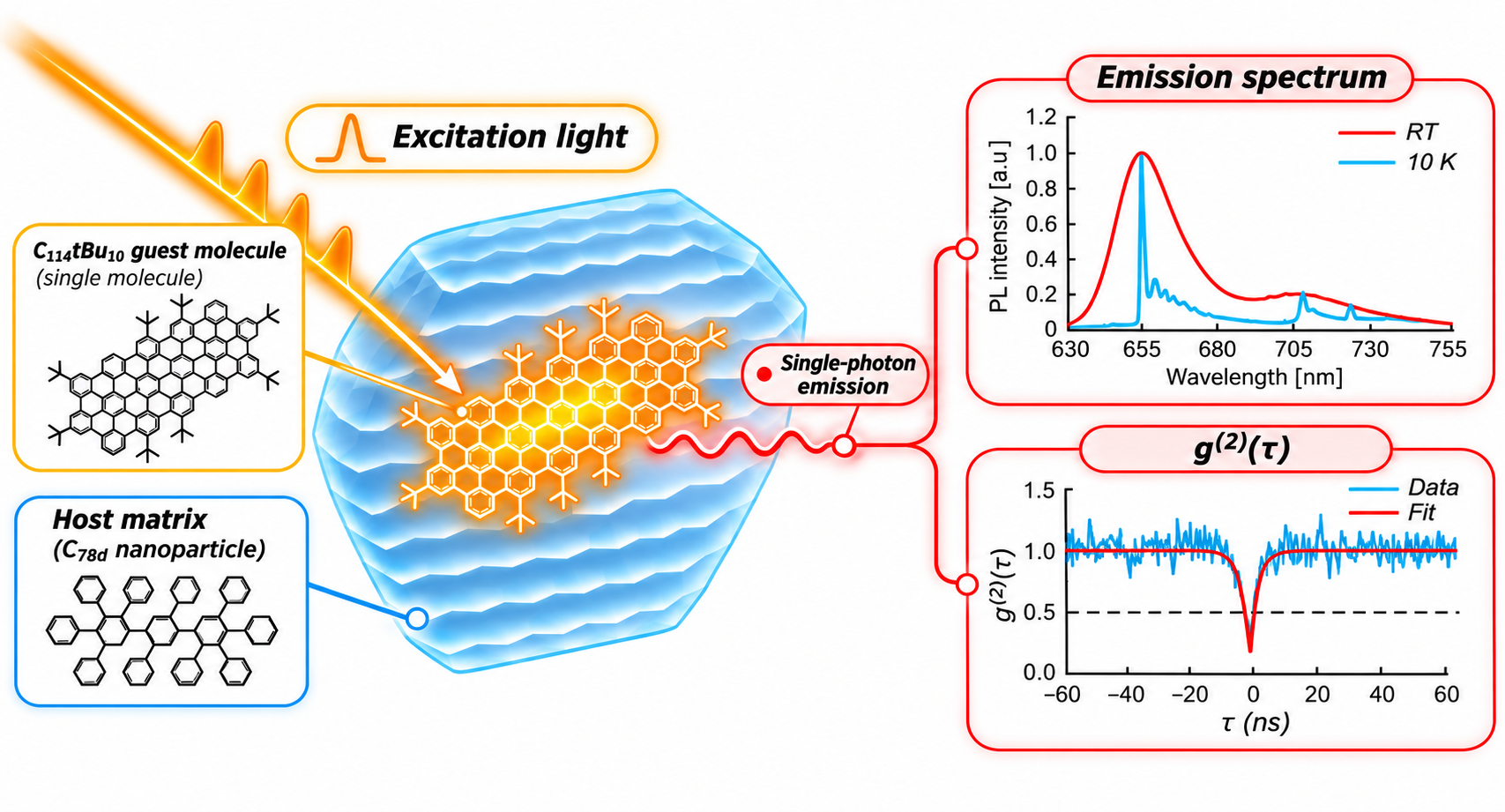}

\end{tocentry}

\begin{abstract}

Graphene quantum dots are promising quantum emitters with stable single-photon emission at room temperature. The next step towards their use in quantum technology applications is to reduce their emission linewidth at low temperatures to reach the radiative limit. To do so, material developments need to involve the control of their local environment. Here, we report a first step in this direction by embedding graphene quantum dots into nanoparticles synthesized from one organic dendrimer precursor. We demonstrate the possibility of isolating a single graphene quantum dot within a single nanoparticle. The photophysics of single graphene quantum dots within isolated nanoparticles is studied at room and cryogenic temperatures. Finally, we show that increasing nanoparticle size reduces spectral jitter and yields linewidths close to the spectral resolution of our experimental setup (hundreds of $\mu$eV).
\end{abstract}

 \begin{figure*}[h!!]
 \centering
 \includegraphics[scale=0.6]{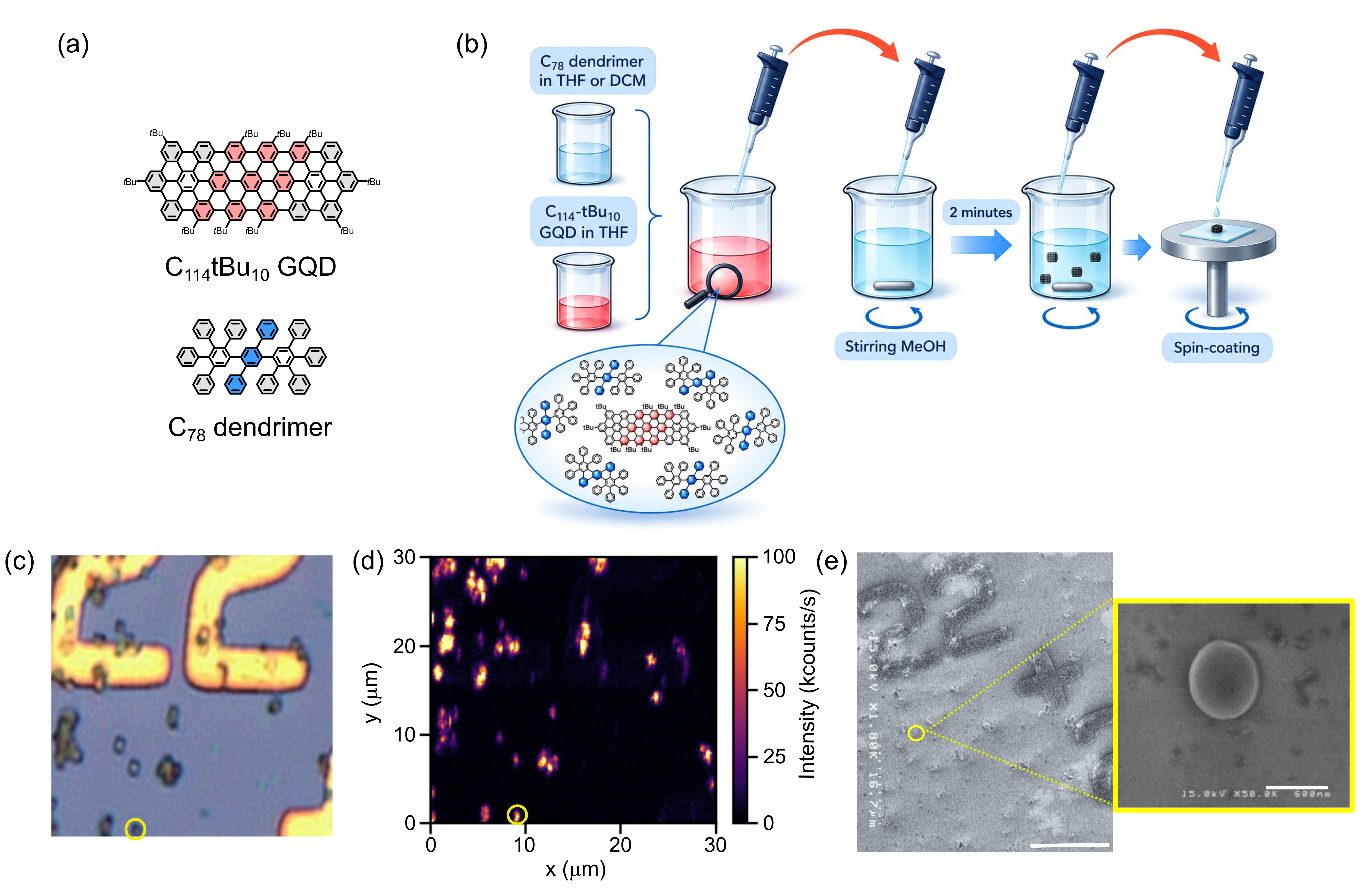}
\caption{(a) Chemical structures of C$_{114}$tBu$_{10}$ GQD and C$_{78}$ dendrimer. (b) Scheme of the synthesis process. (c) Optical image, (d) photoluminescence confocal map, (e) SEM image of the same location on a sample of GQD@C$_{78}$, scale bar is 16.7~$\mu$m. Inset: zoom on the nanoparticle located in the yellow circle, scale bar is 620~nm.}
 \label{cristaux}
\end{figure*}
Molecular quantum nanophotonics is a part of the broad field of quantum technologies. The ability offered by organic chemistry to design, at low cost, a molecule with specific properties is particularly appealing. Nevertheless, only a few molecules in the family of polycyclic aromatic hydrocarbons (PAH) have demonstrated lifetime-limited linewidth at low temperature in specific matrices \cite{toninelli2021single}. This property ensures a long coherence time, which is a prerequisite for most quantum applications. This small number of molecules leaves room for the development of new molecular emitters. Recently, we have reported the synthesis of very large PAHs molecules, i.e. graphene quantum dots (GQD), with excellent optical properties: extremely high fluorescence quantum yield (>90$\%$), short fluorescence radiative life time ($\sim 2-3$ns), high dipole moment (up to 16D) and a size-tunable emission wavelength by more than 100~nm while keeping no size dispersion in the sample \cite{medina2023interplay}. These GQDs show stable single photon emission at room temperature when embedded in a polystyrene matrix \cite{medina2023interplay, levy2023investigation, zhao2018single}. Nevertheless, polymer matrices lead to a broadening of the zero-phonon line (ZPL) at low temperatures, an effect well-reported for various molecules, in particular due to the coupling between the electronic state of the molecule and the low-energy vibration modes of the polymer \cite{kozankiewicz1994single, liu2022vibronic}. This is why specific guest-host systems were developed for small PAH molecules. For instance, Dibenzoterrylene (DBT) molecules in anthracene crystal \cite{lombardi2021triggered} or dibenzanthanthrene (DBATT) in naphtalene~\cite{DBATT1997} show ZPL linewidth limited by the lifetime of the excited state at low temperature. Here, we can note that each guest molecule requires its own specific host.

Inspired by these pioneering works, we report the fabrication of a new molecular system based on organic nanoparticles that enable the inclusion of single 114-carbon-atom GQDs, called C$_{114}$tBu$_{10}$ (see Figure~\ref{cristaux}(a)). In the example of the DBT/anthracene guest/host system, one DBT molecule replaces three anthracene molecules in the molecular crystal \cite{nicolet2007single}. The C$_{114}$tBu$_{10}$ GQD possesses 41 conjugated cycles, whereas DBT has 10. On purely empirical grounds, we choose the C$_{78}$ dendrimer as a building block for the host system (13 cycles, see Fig.~\ref{cristaux}(a)). Indeed, it shares the same rectangular shape as the GQD and maintains the host/guest size ratio. Moreover, the use of organic dendrimers as matrices has already been successful for other single PAH molecules. For instance, pentacene molecules have been embedded into a terphenyl dendrimer matrix, showing lifetime-limited linewidth~\cite{orrit_single_1990}. Finally, in contrast to the PAH molecule, the phenyl groups in the terphenyl and in the C$_{78}$ dendrimer are not conjugated, and their emission is centered in the UV part of the spectrum, far from the emission of the chromophores.

The organic synthesis of the C$_{78}$ dendrimers and  C$_{114}$tBu$_{10}$ has been reported previously \cite{medina2023interplay}. To build C$_{78}$-based molecular nanoparticles, we use a reprecipitation method\cite{THORAT20121}. C$_{78}$ dendrimers are first dissolved in a good solvent (tetrahydrofuran, THF, or dichloromethane, DCM). This solution is then added to an antisolvent (methanol, MeOH) under stirring to induce precipitation. To dope the C$_{78}$ nanoparticles with C$_{114}$tBu$_{10}$, a controlled proportion of well-dispersed GQDs in THF is mixed with the C$_{78}$ solution before the mixture is dropped into the antisolvent. After 2 minutes of stirring, a portion of the solution is extracted and then spin-coated onto a substrate suitable for optical experiments. Finally, the sample is dried in air at room temperature. The process is summarized in Figure~\ref{cristaux}(b).

\begin{figure}[h!]
 \centering
 \includegraphics[scale=.72]{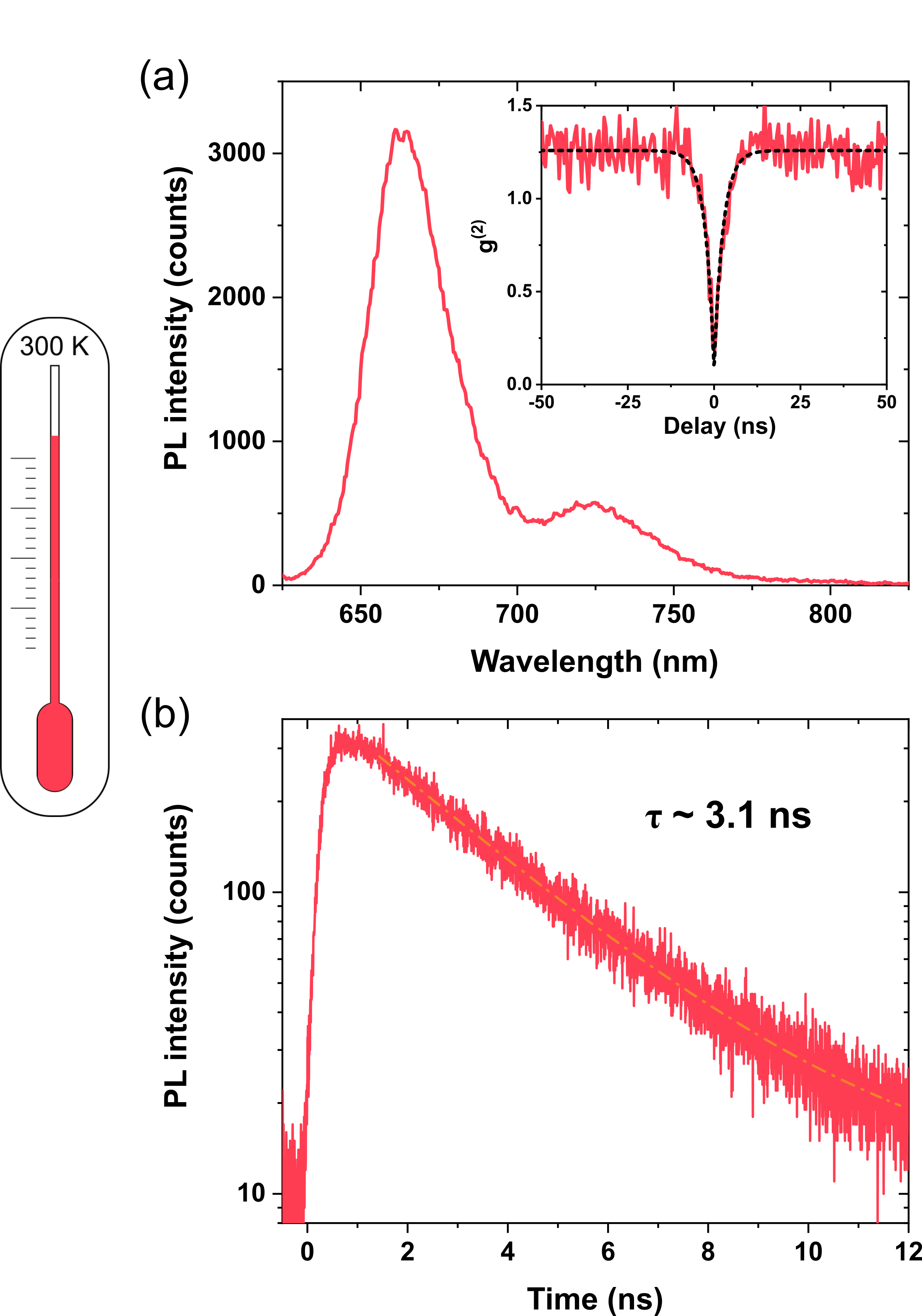}
  \caption{(a) Room temperature photoluminescence spectrum of single C$_{114}$tBu$_{10}$ in a single C$_{78}$ dendrimer nanoparticle. Inset: second-order correlation function as a function of time delay (red), fit with a monoexponential decay ($A-B\exp(-|t|/\tau)$ - black dash line). (b) Typical time-resolved photoluminescence decay at room temperature (red) and fit with a monoexponential decay (orange).}
 \label{roomT}
\end{figure}

The solution of GQD-doped nanoparticles (GQD@C$_{78}$) can be deposited on a Si/SiO$_2$ substrate with markers, thereby allowing localization of the nanoparticles to a specific zone on the sample (see Figure~S1). Figure~\ref{cristaux}(d) and S2 show scanning electron microscope (SEM) images where the particles appear almost spherical, with a diameter of about 600~nm. In some cases, dimers and trimers of fused nanoparticles are observed as well as larger agglomerates (see Figure~S2). A histogram of lateral sizes is displayed in Figure~S8. The nanoparticle mean size is estimated to $\sim$585~nm. In the following, we focus on demonstrating the incorporation of single GQDs into these C$_{78}$ nanoparticles and on studying their photophysics to assess the potential of this new guest-host system.

Figure~\ref{cristaux} shows an optical image (b), a fluorescence confocal map (c), and SEM images(d) of the same location on the Si/SiO$_2$ substrate after the deposition of GQD@C$_{78}$. Nanoparticles and aggregates of nanoparticles are observed in the optical microscope image (see Figure~\ref{cristaux}(c)). The photoluminescence confocal map of the same location (Figure~\ref{cristaux}(d)) shows a direct correlation between the localized nanoparticles and the fluorescence spots. The particle highlighted with the yellow circle in Figures~\ref{cristaux}(c),(d) and (e) shows this spherical shape and emits single photons as demonstrated by second-order correlation measurements (see Figure~S1). Figure~S1 also shows that almost all nanoparticles emit light suggesting that they are all containing GQDs.
Figure~\ref{roomT}(a) displays the spectrum recorded at room temperature on one nanoparticle. It shows the characteristics of the C$_{114}$tBu$_{10}$ GQD, with a zero-phonon line (ZPL) centered at 663~nm and a 0-1 vibronic replica at 726~nm, arising from the C=C stretching mode~\cite{medina2023interplay}. A statistical study on over 70 GQDs reveals a ZPL line centered at $666\pm 3.7$~nm. The mean value of the ZPL wavelength is comparable to the one obtained in a polystyrene matrix (664~nm). This indicates that the C$_{78}$ nanoparticle environment barely modifies the GQD optical transitions. The second-order correlation measurements depicted in the inset of Figure~\ref{roomT}(a) show the characteristic anti-bunching behavior with a g$^2$(0)$\sim$0.1<0.5, demonstrating that a single GQD is embedded in this C$_{78}$ nanoparticle.  Figure~S4 shows a statistic of g$^{(2)}$ at zero delay recorded on 67 spots. More than 80$\%$ of the spots show g$^2(0)<0.5$. This demonstrates the process's reliability in incorporating a single GQD into a single C$_{78}$ nanoparticle. Moreover, the luminescence can remain stable for hundreds of seconds under ambient conditions (see Figure~S5). The statistics of fluorescence stability show that over 60$\%$ of spots live longer than 300s for a laser excitation intensity of $\sim$0.8~$kW.cm^{-2}$ in ambient conditions. Moreover, the emission of a single GQD in the C$_{78}$ nanoparticle shows a strong dependence of the fluorescence signal as a function of the incident polarization that corresponds to the dipole aligned with the main axis of the GQD~\cite{medina2023interplay}. The curves are well fit by a Malus law, with an extinction coefficient ranging from 84$\%$ to 93$\%$ (see Figure~S6). Finally, a typical time-resolved fluorescence decay is plotted in Figure~\ref{roomT}. The mono-exponential fit gives a time decay of $\tau\sim 3.1$ns. Statistics for 18 single GQDs yield an average lifetime of 3.2~ns with a standard deviation of 0.6~ns (see Figure~S4). We discuss these values in terms of changes in radiative lifetime and modifications of nonradiative recombination channels below.
\begin{figure}[h!]
 \centering
 \includegraphics[scale=0.72]{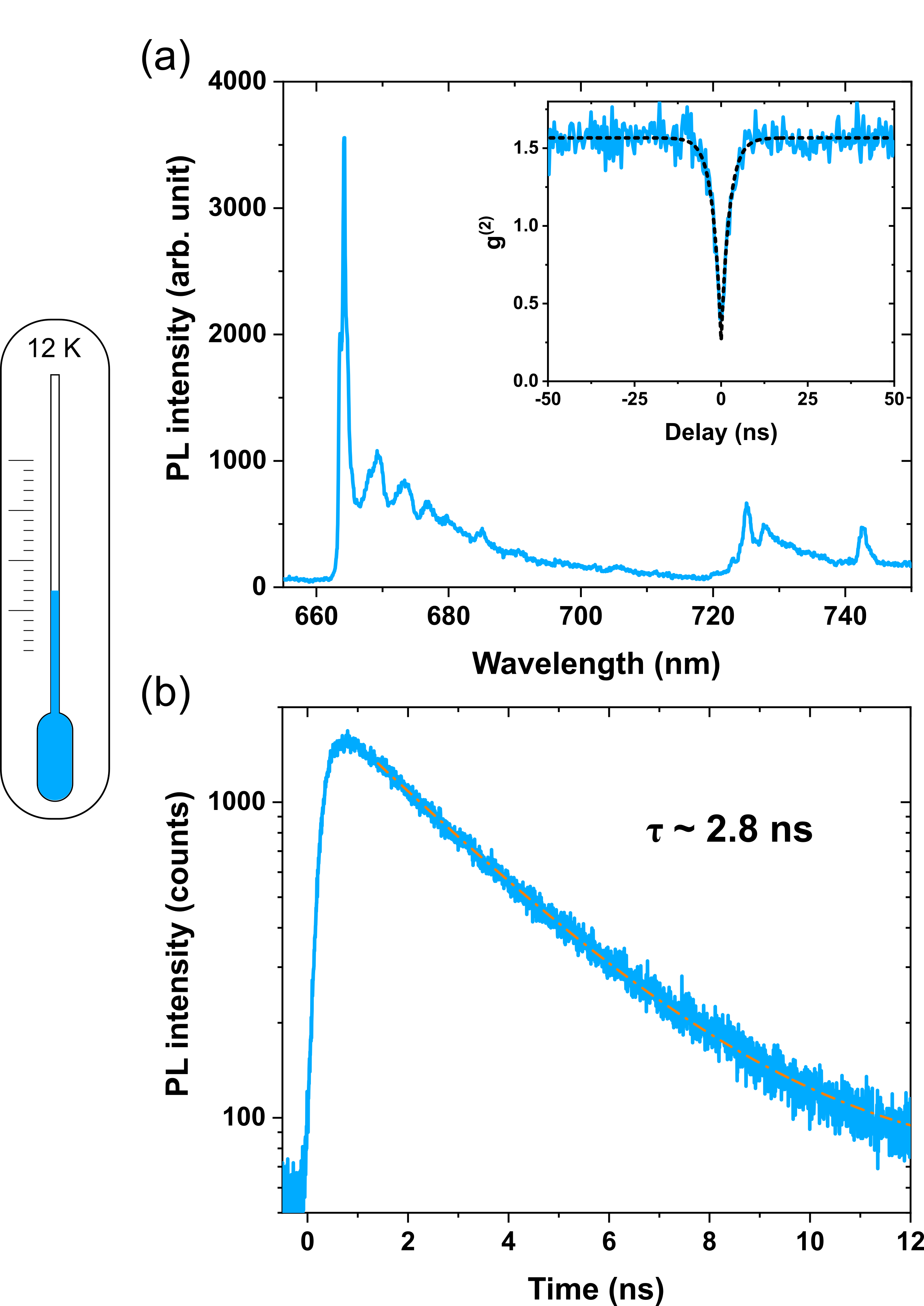}
\caption{(a) Low temperature photoluminescence spectrum of a single C$_{114}$tBu$_{10}$ in a single C$_{78}$ nanoparticle ($\sim$12K). Inset: second-order correlation function measured for the same spot, fit with a monoexponential decay ($A-B\exp(-|t|/\tau)$ - black dash line). (b) Typical time-resolved photoluminescence decay recorded at low temperature (blue) and fit with a monoexponential decay (orange).}
 \label{lowT}
\end{figure}

In solution (THF), the lifetime is 2.6~ns, and we measured quantum yield above 90$\%$. This leads to a radiative lifetime of $\sim$2.89~ns. For comparison, single GQDs in polystyrene show an average lifetime of $\sim$2.3~ns with a standard deviation of $\sim$0.1~ns. In that case, the main effects that can account for this decrease in the total lifetime are the modification of the dielectric environment and the decrease in the fluorescence quantum yield due to the activation of non-radiative decay channels. The modification of the lifetime induced by the dielectric environment is a complex phenomenon. One rough way to account for the variation of radiative lifetime with the refractive index is to take $\tau_R(n)=\frac{\tau_{vac}}{n}$ where $\tau_{vac}$ is the radiative lifetime in vacuum \cite{he_localfield_2010}. Using the refractive indices of THF (1.404) and polystyrene (1.580) yields $\tau_{R; PS}\sim2.57$~ns for the GQD in polystyrene. The measured total lifetimes in PS range from 2.1~ns to 2.55~ns, yielding a fluorescence quantum yield of 82$\%$ to 99$\%$. We conclude that the lifetime modification is primarily explained by a change in the refractive index, with no significant change in the PL quantum yield. The refractive index of the C$_{78}$ nanoparticles is unknown. Nevertheless, because the average ZPL wavelength is very close to the one measured in the polystyrene matrix, one can argue that the local refractive index should not differ much. Yet, the average value of the total lifetime is significantly longer ($\tau_{PS}\sim 2.3$~ns; $\tau_{C78}\sim 3.2$~ns). While non-radiative decay channels shorten the total lifetime, the higher mean value measured here shows that the nanoparticle environment maintains the GQD's high fluorescence quantum yield. Likewise, it indicates that the radiative lifetime is the one that is modified. Interestingly, a large portion of the measured lifetimes is even longer than the expected radiative lifetime. Qualitatively, it suggests that other processes might modify the radiative decay of the GQD in the C$_{78}$ nanoparticle, since we do not expect a large void (smaller refractive index toward n=1) around the GQDs inside the nanoparticle. While a modification of the steric environment might modify the rigidity of the GQD structure and lead to a change in the oscillator strength, another explanation would be the proximity of the GQDs to the C$_{78}$/air interface. In particular, it has been shown that when terrylene molecules are embedded in a para-terphenyl crystal of $\sim$20~nm thickness, their radiative rate decreases as they approach the crystal surface and is a few times lower than the one in the bulk crystal \cite{chen_enhancing_2015}. In this context, the median of the total lifetime (3.16~ns$> \tau_{R; PS} \sim\tau_{R; C78} \sim 2.57$ns) suggests that more than 50$\%$ of the GQDs would be close enough to the surface to see their radiative lifetime increased (see Figure S4).

\begin{figure*}[h!]
 \centering
 \includegraphics[scale=0.75]{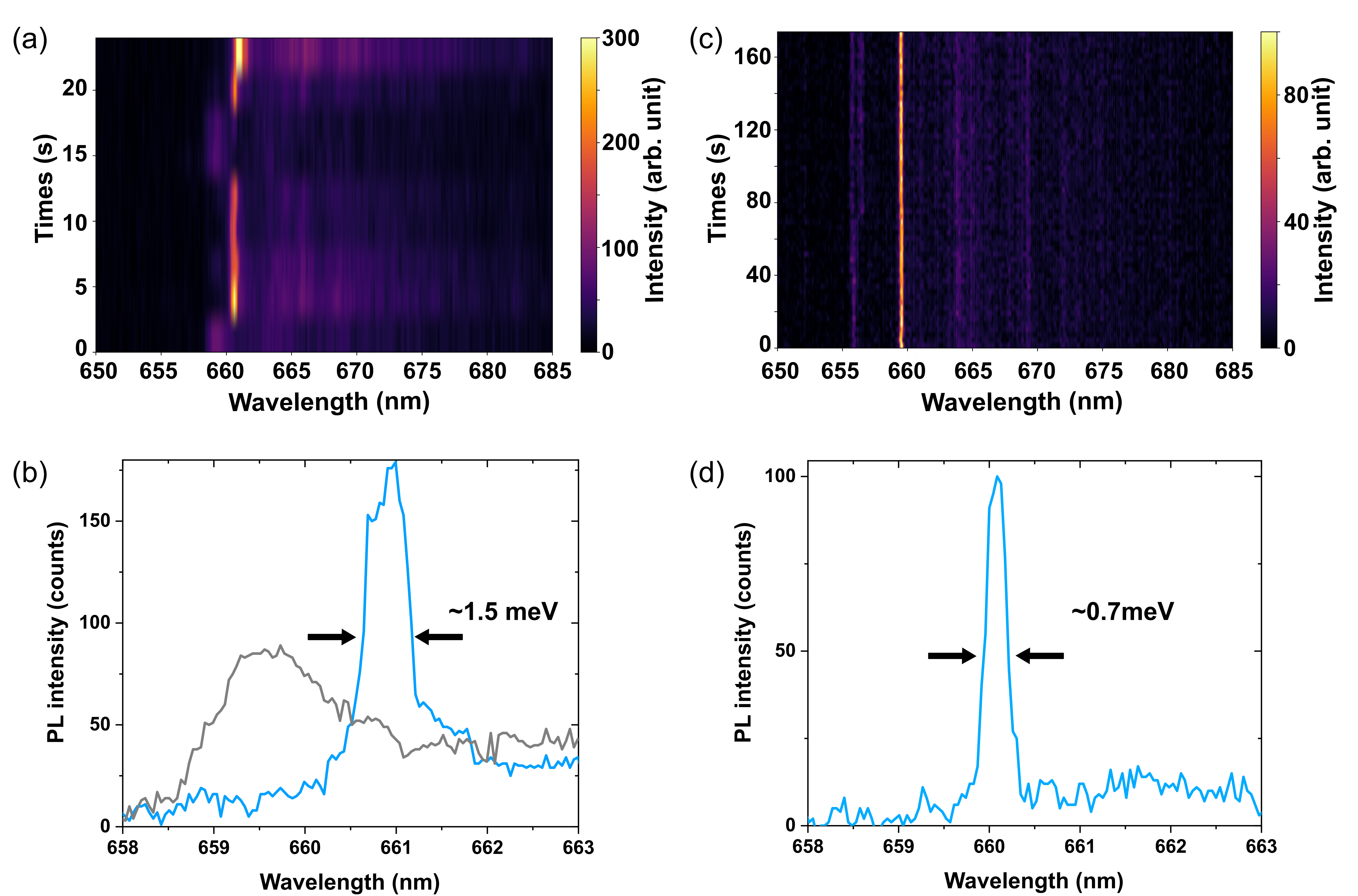}
\caption{(a) Photoluminescence spectra as a function of time at 12~K recorded a standard nanoparticle. Time acquisition: 3~s. Excitation intensity 13kW.cm$^{-2}$ (b) Two typical spectra extracted from (a).(c) Photoluminescence spectra as a function of time at 12~K recorded on a larger nanoparticle. Time acquisition: 1~s. Excitation intensity 13kW.cm$^{-2}$ (d) one typical spectrum extracted from (c).}
 \label{jitter}
\end{figure*}

In the previous section, we demonstrated that these new molecular nanoparticles can host single GQDs with a high brightness. We then conducted experiments at cryogenic temperatures to reduce coupling to vibrational modes and to asses their potential in terms of temporal coherence. Figure~\ref{lowT} shows the fluorescence spectrum of a GQD at $\sim 12$~K. The $g^{(2)}(\tau)$ in the inset of Figure~\ref{lowT} confirms that a single GQD is under study. The linewidth of the zero-phonon line (ZPL) is reduced to a few meV, and vibrational replicas are resolved both at low (hundreds of $cm^{-1}$) and high (1300-1600~$cm^{-1}$) frequencies. Interestingly, when the spectra recorded in the nanoparticle and in the polystyrene are superimposed (see Figure~S7), low-frequency vibrational modes match very well, again showing that the molecular nanoparticles preserve the main GQDs properties even at low temperature.  Moreover, a zoom on the ZPL (see Figure~S7) shows that the line is asymmetric. As we will see later, this observation is related to the spectral diffusion of the line during the acquisition time window. Figure~\ref{lowT}(b) also displays a typical TRPL decay of a single GQD at cryogenic temperatures. It remains monoexponential. We recorded TRPL decay times for four single GQDs, and the extracted lifetimes range from 2.7~ns to 3.27~ns, which are comparable to the room-temperature results. The stability of the lifetime value as temperature decreases indicates that non-radiative channels must be negligible at room and cryogenic temperatures, with the fluorescence quantum yield remaining close to unity.

As mentioned previously, the ZPL line shape is not symmetric, and the PL linewidth is still high for experiments performed at $\sim 12$~K. In order to investigate these phenomena, the acquisition time has been decreased from 30~s to 3~s and spectra as a function of time have been recorded (see Figure\ref{jitter}(a)). It is clear that the lines show spectral jumps over time, as well as intensity fluctuations. Increasing the acquisition time averages the spectrum over these different positions, resulting in an apparent broadening. Figure~\ref{jitter}(b) shows two typical spectra, extracted from the time-trace, centered at different wavelengths.  The reddest one shows a reduced linewidth down to 1.5~meV. Most likely, faster fluctuations may occur within the 3~s acquisition time, leading to inhomogeneous broadening. Spectral jittering on single quantum emitters at cryogenic temperatures is commonly observed, especially when they are placed near a surface. Stark effect, due to charge fluctuations near the emitter, is known to be the main cause of spectral and intensity jittering \cite{orrit1992stark,reigue2019resonance}. For instance, few meV spectral jumps has been reported for the luminescence of carbon nanotubes at 4.2~K due to electrostatic fluctuations on the SiO$_2$ substrate \cite{Noe2018}. Likewise, NV centers located tens of nanometers from the surface show spectral jumps of tens of GHz (tens to hundreds of $\mu$eV) due to surface charge fluctuations \cite{NV}. Furthermore, it has been shown in other guest/host systems that the inhomogeneous broadening due to spectral jumps depends on the thickness of the host crystal \cite{Gmeiner2016}. In this last example, the increase in defect density, acting as fluctuating two-level systems, while decreasing the crystal size, would explain the GHz spectral fluctuations, even though the influence of surface charge fluctuations cannot be entirely excluded. In the example plotted in Figure~\ref{jitter}, the amplitude of the jitter is of the order of 5~meV. It suggests a highly fluctuating environment of the single GQD. As in the examples mentioned above, the jitter may be due to the proximity of the GQD to the nanoparticle surface, consistent with our interpretation of the increase in fluorescence lifetime discussed in the first part of this paper. Likewise, poor crystallinity in these small nanoparticles might also explain the large spectral fluctuations. To take things a step further, we decided to increase the nanoparticle size to increase the probability that the GQD would be farther from the surface and, eventually, to decrease the density of defects. This was done by increasing the starting concentration of C$_{78}$ by a factor of two, leading to a mean lateral size of $\sim$900~nm whereas it was 585~nm for the standard nanoparticles (see Figure~S8). Figure~\ref{jitter}(c) shows the time trace of a GQD at a low temperature on a single nanoparticle made with this preparation method. The emission is much more stable in terms of intensity and spectral jitter. The measured linewidth within 1~s integration time decreases to $\sim$0.7~meV, which corresponds to the current spectral resolution of our setup. These encouraging preliminary results open the way to further improvements. First, statistical measurements are now needed to link nanoparticle size to GQD emission stability. Secondly, structural investigations using transmission electron microscopy are planned to elucidate the crystallinity of the nanoparticles as a function of the experimental growth conditions.

\section{Conclusion}
 In conclusion, we have reported the synthesis of a new guest/host system based on C$_{114}$tBu$_{10}$ GQDs and a C$_{78}$ dendrimer nanoparticle. The developed synthesis enables the isolation of single GQDs within the nanoparticle, as demonstrated by the g$^2(\tau)$ measurements. The high fluorescence quantum yield of the GQD in solution is preserved in the solid-state C$_{78}$ dendrimer nanoparticle, and a narrow linewidth limited by the current resolution of our setup has been observed in the case of larger nanoparticles. Improving the synthesis by embedding the GQDs deeper away from the nanoparticle surface and increasing their crystallinity will lead to a significant reduction in the linewidth by decreasing spectral jittering, as suggested by our preliminary results. Moreover, the low-temperature experiments reported here have been limited to 12~K. Lowering the temperature to a few Kelvin may also reduce the linewidth. For comparison, a DBATT single molecule in naphthalene shows more than one order of magnitude reduction in linewidth between 11K and 3K~\cite{tamarat_ten_2000}. Finally, a photoluminescence excitation experiment with sub-MHz resolution is currently being implemented on our experimental setup to measure linewidths close to the lifetime limit.



\begin{suppinfo}

\end{suppinfo}

\begin{acknowledgement}
This work was ﬁnancially supported by the ANR grant DELICACY ANR-22-CE47-0001-03, ANR grant GANESH ANR-21-CE09-0025. This work has been supported as part of France 2030 programme ANR-11-IDEX-0003. This work received government funding under the France 2030 programme (QuanTEdu-France), reference number ANR-22-CMAS-0001
\end{acknowledgement}

\bibliography{biblio_C78d2}

\newpage
\onecolumn
\setcounter{figure}{0}
\renewcommand{\thefigure}{S\arabic{figure}}

\begin{center}
\Large \textbf{Supplementary information - Single Graphene Quantum dots in a single molecular nanoparticle}\\
\end{center}
\textbf{Co-localization of C$_{78}$ nanoparticles doped with a single GQD}\\
 \begin{figure}[h!]
 \centering
 \includegraphics[scale=.75]{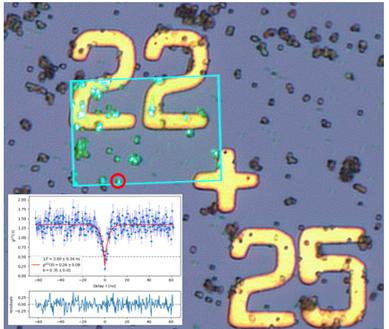}
\caption{Optical microscope image of C$_{78}$ nanoparticles superimposed with fluorescence map of the same zone (blue rectangle). Inset: g$^{2}(\tau)$ function recorded on the nanoparticle highlighted in red showing an antibunching at zero delay.}
 \label{coloc}
\end{figure}

\vspace{3cm}

\textbf{SEM images}

 \begin{figure}[h!!]
 \centering
 \includegraphics[scale=0.75]{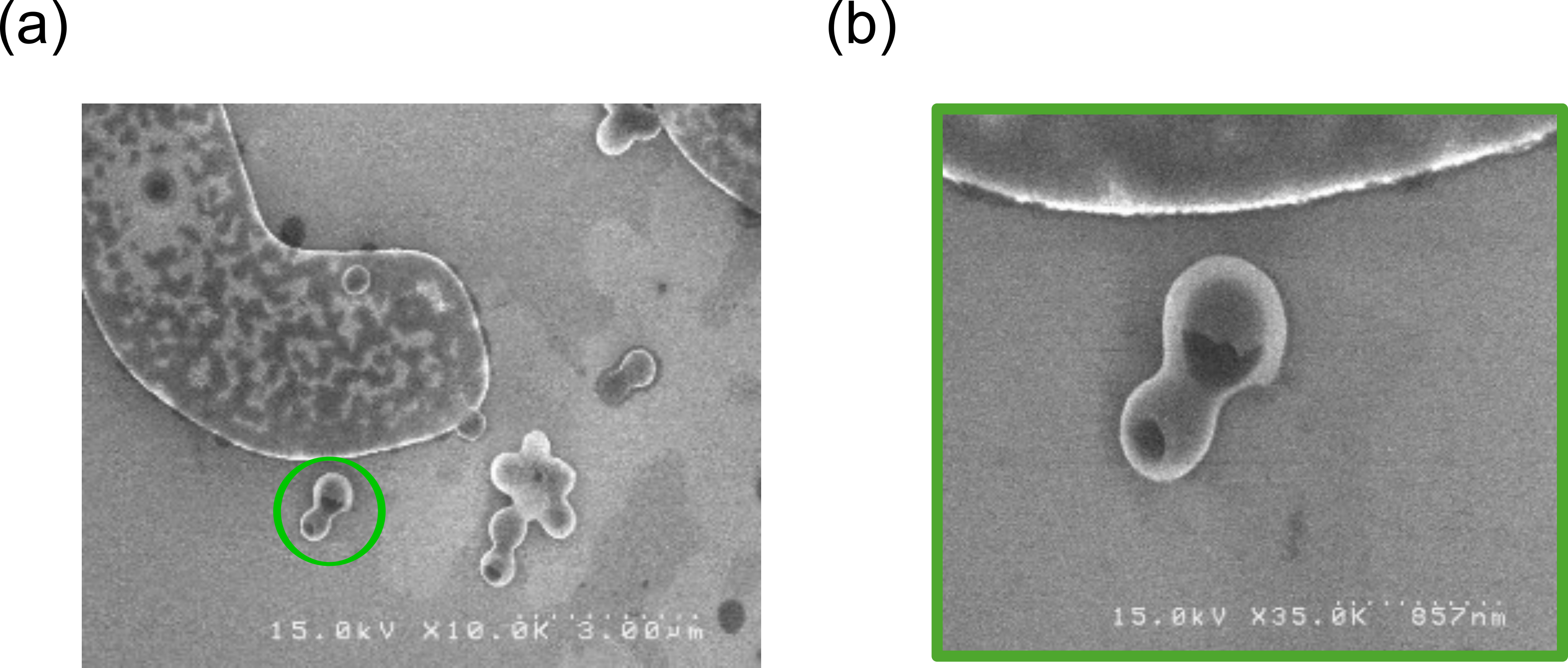}
\caption{SEM images of C$_{78}$ nanoparticles doped with C$_{114}$tBu$_{10}$ GQD. (b) zoom on the zone highlighted in green in (a).}
 \label{caracterization}
\end{figure}

\newpage

\textbf{Confocal fluorescence maps of C$_{78}$ nanoparticles doped with a high concentration of GQDs}\\

 \begin{figure}[h!]
 \centering
 \includegraphics[scale=0.25]{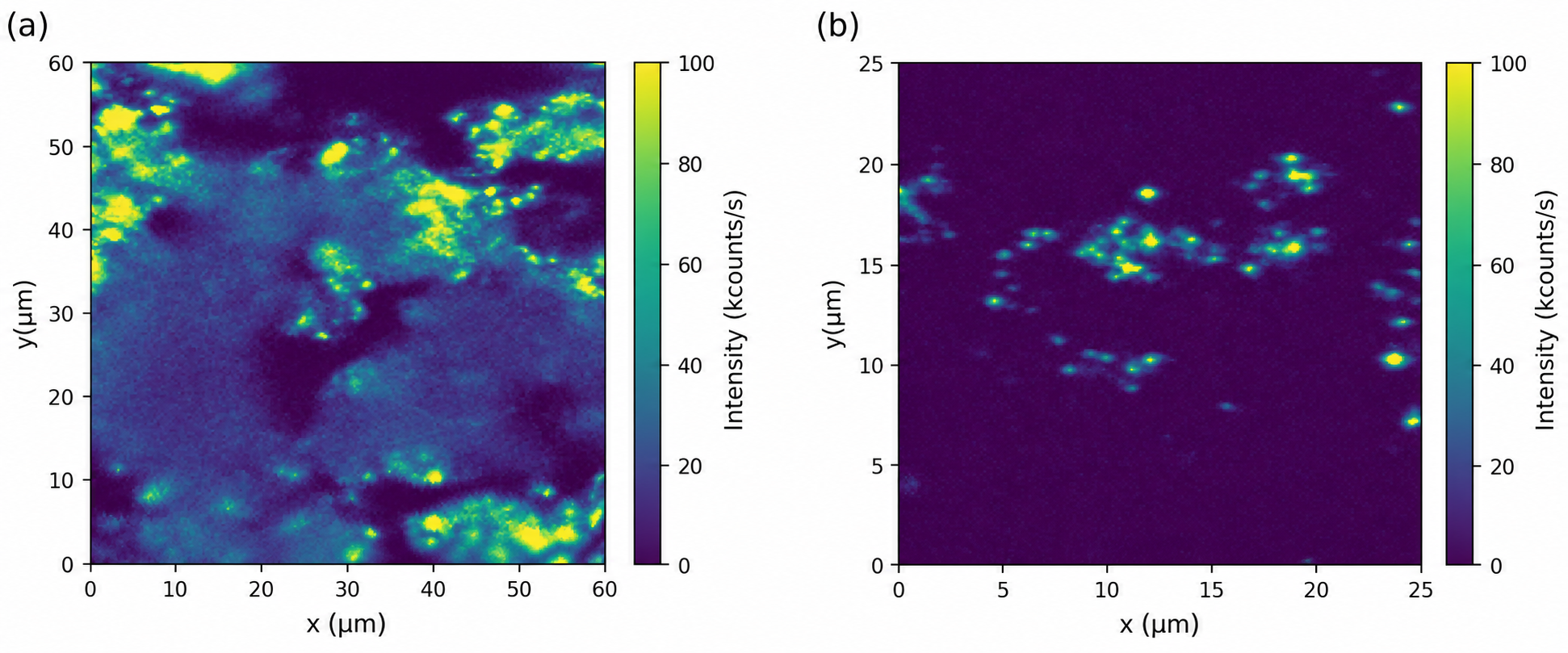}
\caption{Confocal maps for different concentration of GQDs. (a) Solution of GQD with OD of 1e$^{-4}$ in OD. (c) Solution of GQD with OD 6e$^{-6}$}
 \label{HighC}
\end{figure}

\vspace{3cm}

\textbf{Second order correlation measurements}\\
 \begin{figure}[h!]
 \centering
 \includegraphics[scale=.3]{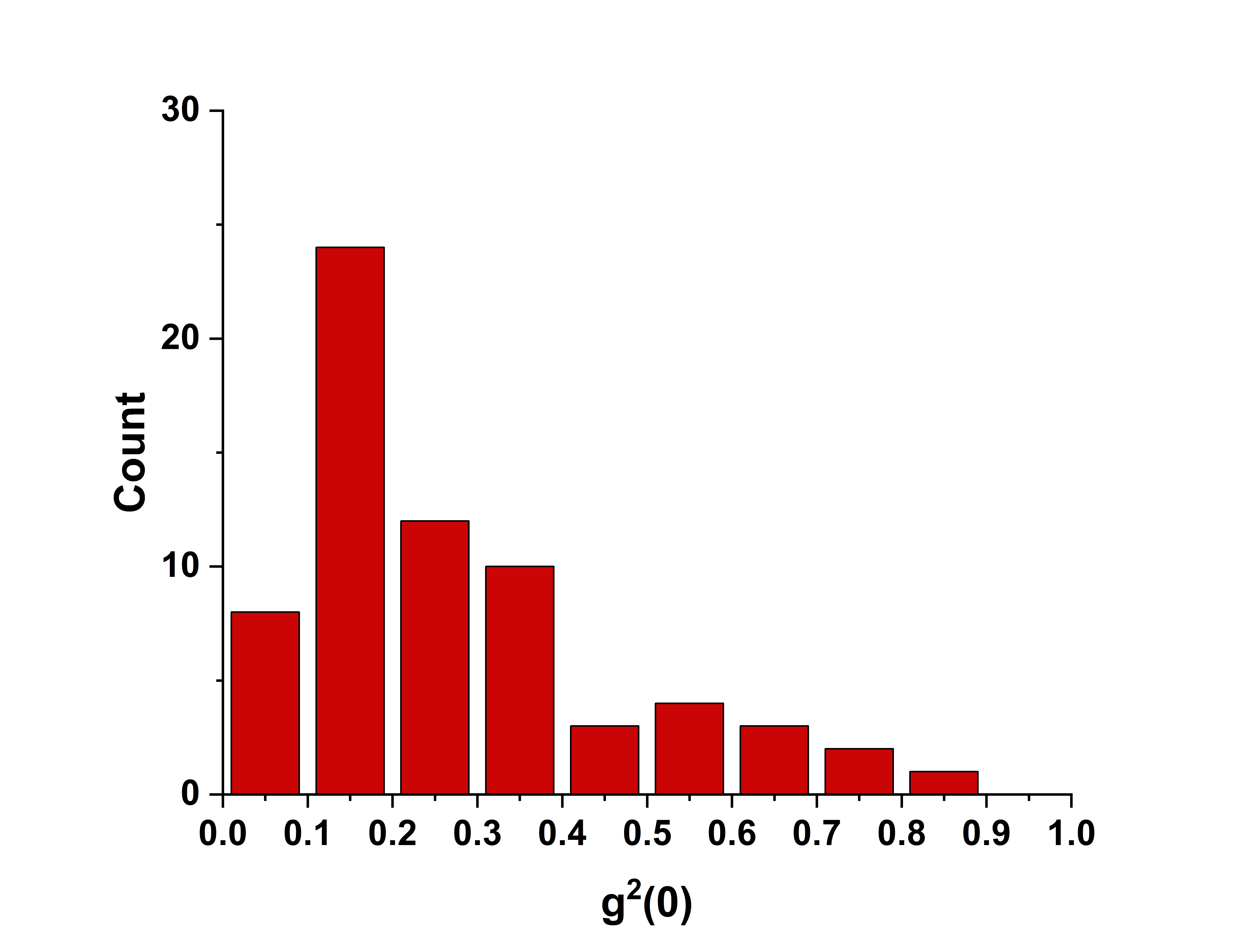}
  \includegraphics[scale=.3]{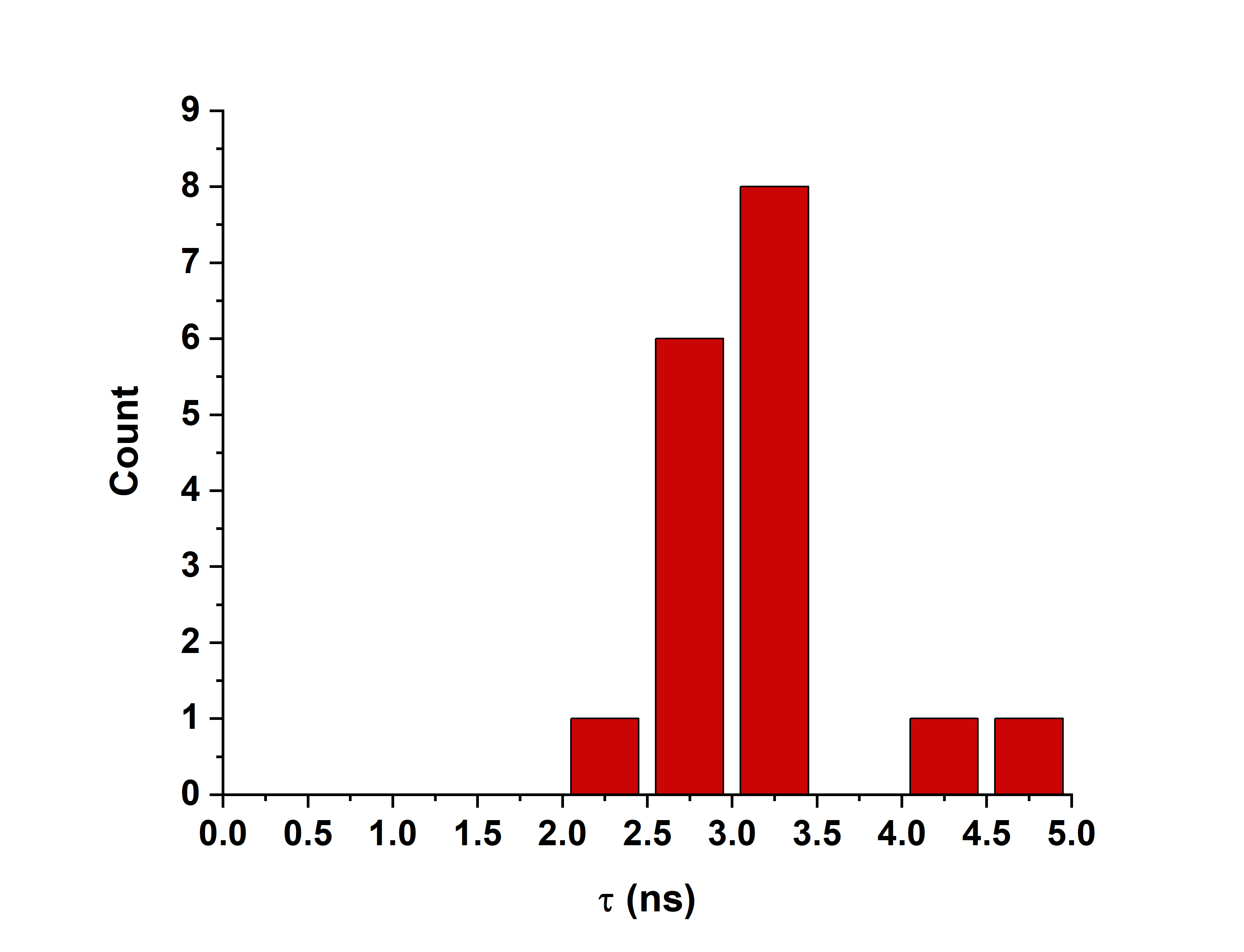}
\caption{(Left panel) Values of the g$^{(2)}(0)$ recorded on 67 GQD spots on the same sample. (Right panel) Statistics of the life time measured on single GQD in C$_{78}$ nanoparticles. }
 \label{statg2}
\end{figure}
\vspace{2cm}

\textbf{Photostability}\\
 \begin{figure}
 \centering
 \includegraphics[scale=.15]{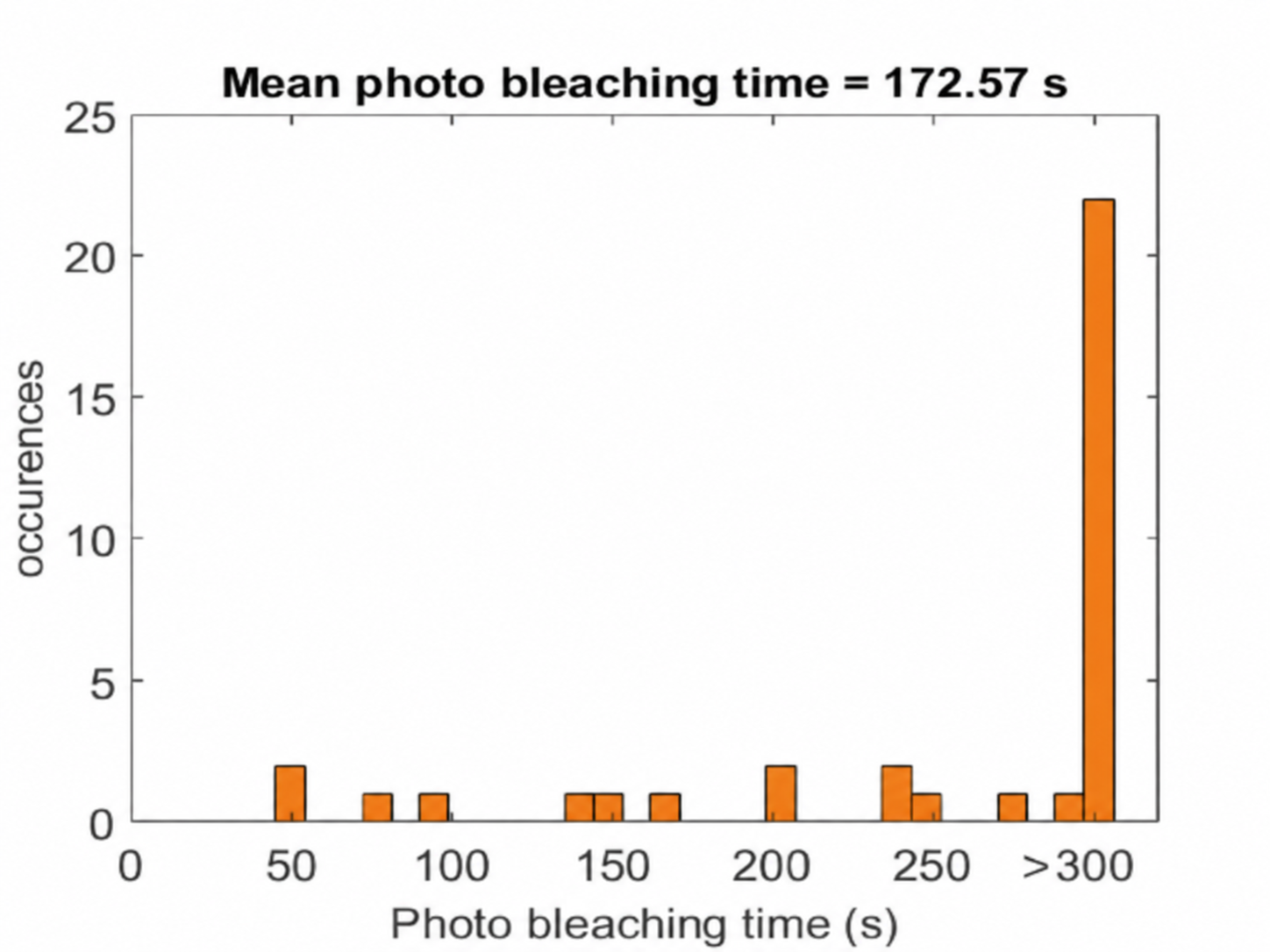}
\caption{Statistics of the photostability of single GQDs in C$_{78}$ nanoparticles using a low excitation intensity $\sim 0.78 kW.cm^{-2}$. }
 \label{photstab}
\end{figure}

\textbf{Polarization measurements}\\
 \begin{figure}[h!]
 \centering
 \includegraphics[scale=.5]{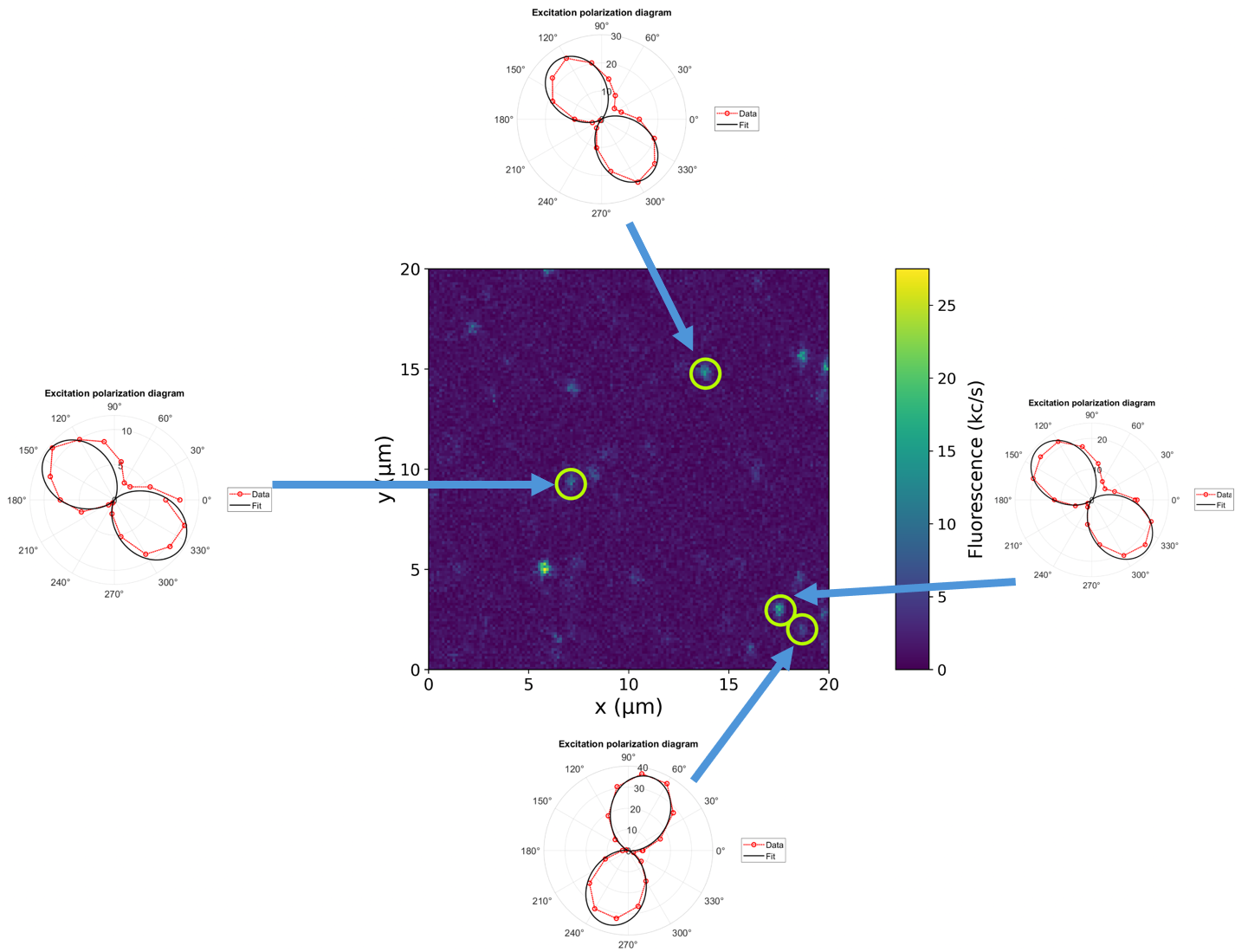}
\caption{Examples of confocal map together with polarization resolved measurements.}
 \label{polar}
\end{figure}
\newpage

\textbf{Low temperature measurements}\\
 \begin{figure}[h!]
 \centering
 \includegraphics[scale=.45]{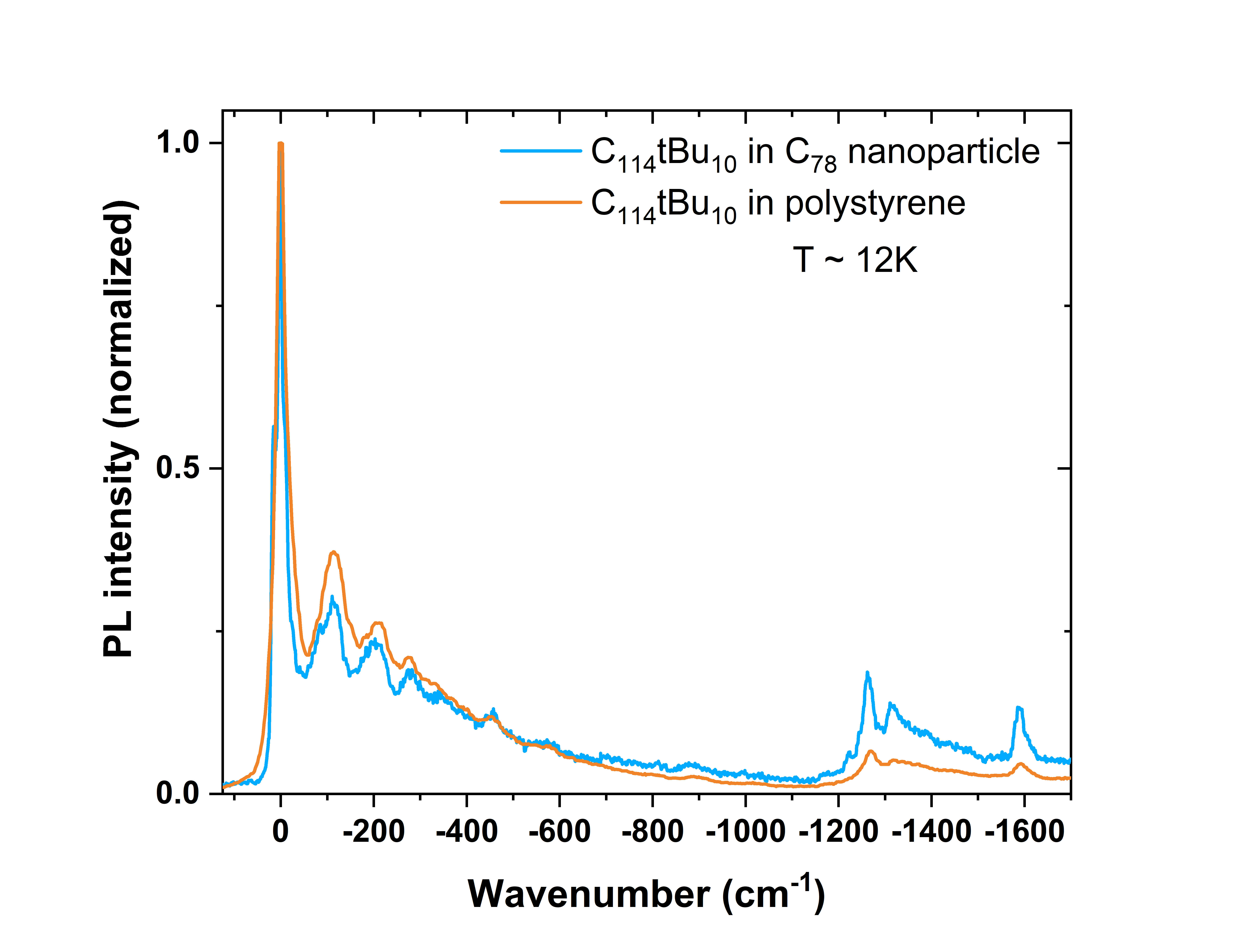}
  \includegraphics[scale=.45]{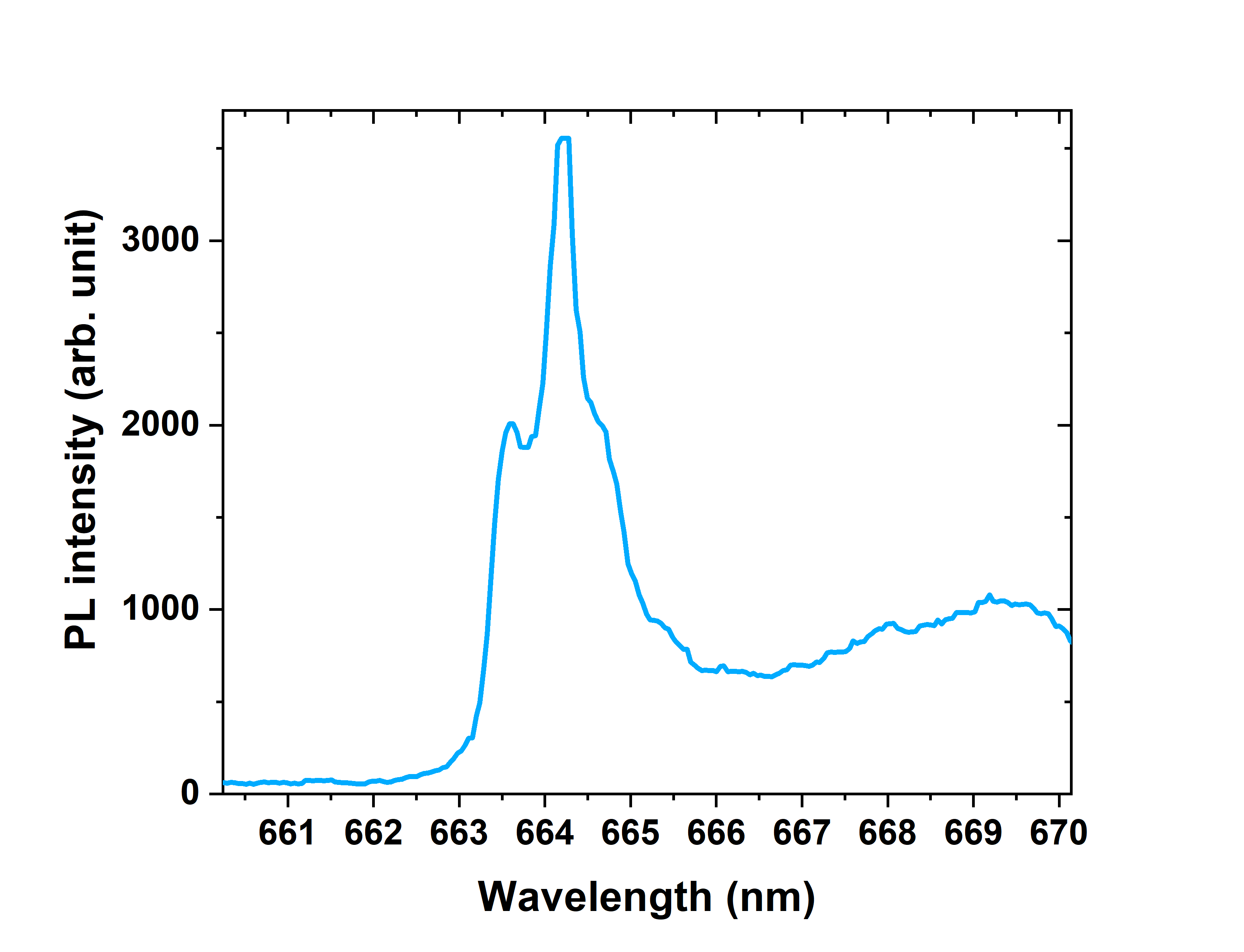}
\caption{(top panel) Comparison of low temperature spectra of a single GQD embedded in polystyrene (orange) and in a in C$_{78}$ nanoparticle ; (Bottom panel ) Zoom on the zero-phonon line of the low T spectrum reported in main Figure~3.}
 \label{LowTps}
\end{figure}

\textbf{Increasing the size of the nanoparticle}\\
All the experiments showed in the main paper has been done for starting concentration of C$_{78}$ of 1mg/mL. In order to increase the average size of the nanoparticle, the concentration has been increased to 2mg/mL. SEM measurements has been performed on Si/SiO$_2$ substrate. The starting solution at 0.1mg/mL leads to "small" crystals with a mean size of $\sim$585~nm, and a median at $\sim$590~nm. At 0.2mg/mL bigger crystals are obtained with a mean size of $\sim$900~nm, median $\sim$870~nm.

 \begin{figure}
 \centering
 \includegraphics[scale=.45]{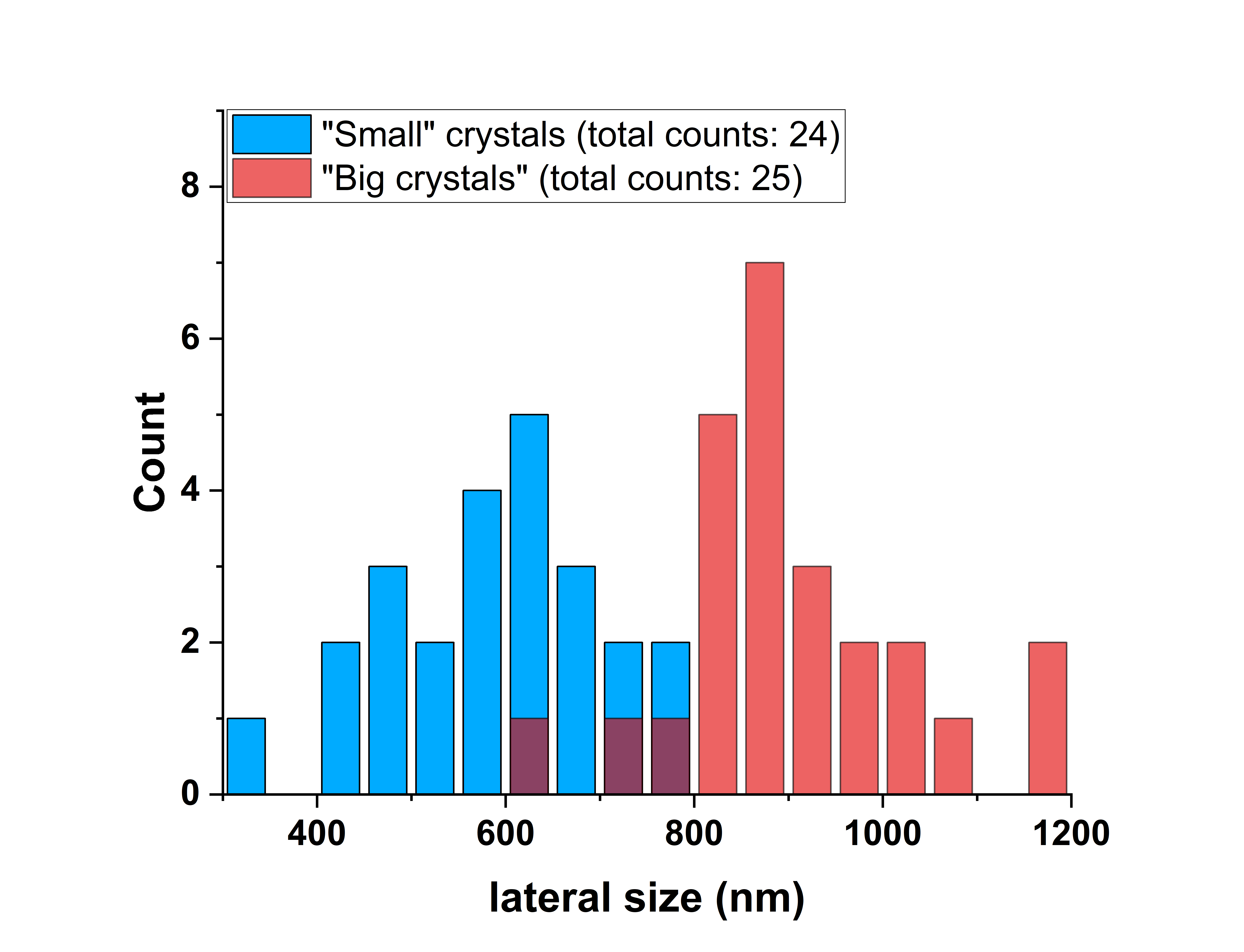}
\caption{Histogam of nanoparticle lateral size for starting concentration of C$_{78}$ of 1mg/mL (blue) and  2mg/mL (red).}
 \label{histo_size}
\end{figure}

\end{document}